\documentstyle[12pt]{article}

\def\seCtion#1{\section{#1} \setcounter{equation}{0}}
\renewcommand\theequation{\ifnum\value{section}>0{\thesection.
\arabic{equation}}\fi}
\newcommand{\be}{\begin{equation}}
\newcommand{\ee}{\end{equation}}
\newcommand{\bea}{\begin{eqnarray}}
\newcommand{\eea}{\end{eqnarray}}
\newcommand{\nn}{\nonumber}
\newcommand{\bi}{\bibitem}
\newcommand{\la}{\label}

\begin{document}

\pagestyle{empty}
\begin{flushright}
FIUN-CP-01/1
\end{flushright}

\begin{center}
{\Large \bf Rest masses of elementary particles as \\
effective masses at zero temperature}\\

\vspace{0.3cm}
{\bf C. Quimbay \footnote{Associate researcher of CIF, Bogot\'a, 
Colombia. carloqui@ciencias.unal.edu.co } and 
J. Morales \footnote{Associate researcher of CIF, Bogot\'a, 
Colombia. johnmo@ciencias.unal.edu.co}}\\
{\it{Departamento de F\'{\i}sica}, Universidad Nacional de 
Colombia\\
Ciudad Universitaria, Bogot\'a, Colombia}\\
\vspace{0.5cm}
October 3, 2001
\end{center}

\vspace{0.8cm}

\begin{abstract}
We introduce a new approach to generate dinamically the masses of 
elementary particles in the $SU(3)_C \times SU(2)_L \times U(1)_Y$ 
Standard Model without Higgs Sector (SMWHS). We start from the 
assumption that rest masses correspond to the effective masses 
of particles in an elementary quantum fluid at zero temperature.
These effective masses are obtained through radiative corrections, 
at one-loop order, in the context of the real time formalism of 
quantum field theory at finite temperature and density. The quantum 
fluid is described in structure and dynamics by the SMWHS and it 
is characterized by non-vanishing chemical potentials associated 
to the different fermion flavour species. Starting from the 
experimental mass values for quarks and leptons, taking the 
top quark mass as $m_t = 172.916$ GeV, we can compute, as an 
evidence of the consistency of our approach, the experimental 
central mass values for the $W^{\pm}$ and $Z^0$ gauge bosons. 
Subsequently we introduce in the SMWHS a massless scalar 
field leading to Yukawa coupling terms in the Lagrangian 
density. For this case we can also compute the experimental 
mass central values of the $W^{\pm}$ and $Z^0$ gauge bosons 
using a top quark mass value in the range $169.2$ GeV 
$< m_{t} < 178.6$ GeV; this range for the top quark mass 
implies that the scalar boson mass must be in the range 
$0 < M_{H} < 152$ GeV.
\end{abstract}

\newpage

\pagestyle{plain}

%%%%%%%%%%%%%%%%%%%%%%%%%%%%%%%%%%%%%%%%%%%%%%%%%%%%%%%%%%%%%%%%%%%%%%

\seCtion{Introduction}

\hspace{3.0mm}
The determination of the mechanism by which elementary particles 
acquire mass is an important step in the aim to understand nature 
at the microscopic level \cite{gunion}. In this determination the 
Higgs mechanism is one of the most atractive approaches known in 
particle physics \cite{gunion}. However this mechanism, which is 
based on the existence of a Higgs sector of scalar fields, is the 
most intringuing pillar of the Standard Model (SM) \cite{sirlin}. 
The spin-zero Higgs field is a doublet in the SU(2) space carrying 
non-zero hypercharge, and it is a singlet in the SU(3) space of 
color. The Higgs potential must be such that one of the neutral 
components of the Higgs field spontaneously acquires a 
non-vanishing vacuum expectation value, thereby giving masses to 
the $W^{\pm}$ and $Z^0$ gauge bosons \cite{quigg, altarelli}. In 
this way the $SU(2)_L \times U(1)_Y$ symmetry is spontaneously 
broken into the $U(1)_{em}$ symmetry. Simultaneously the Yukawa 
couplings between the Higgs boson and fermion fields lead to the 
generation of masses for the elementary fermions. As a consequence, 
the Lagrangian is symmetric under $SU(2)_L \times U(1)_Y$ 
transformations but the vacuum is not \cite{quigg}. Bosonic gauge 
fields and fermionic matter fields acquire masses through their 
interactions with the Higgs field. The above mechanism implies 
the existence of a new particle in the physical spectrum, the 
neutral Higgs boson. The search for the Higgs boson is a focus of 
present \cite{degrassi}-\cite{ridolfi} and future experimental 
research \cite{gunion}. The possible existence of the Higgs boson 
with a mass $M_H \approx 115$ GeV, of which there is no conclusive 
evidence from LEP, must be confirmed or discarded at LHC in the 
future \cite{gunion}. On the other hand, it is important to 
show up that the couplings of quarks and leptons to the weak gauge 
bosons, $W^{\pm}$ and $Z^0$, are indeed precisely those prescribed 
by the electroweak gauge symmetry of the SM \cite{altarelli}. The 
triple gauge vertices $\gamma W^+ W^-$ and $Z^0 W^+ W^-$ have also 
been found in agreement with the specific prediction of the SM at 
tree level \cite{altarelli}. These facts have been proved with 
high accuracy through the experimental precision tests of the SM 
\cite{altarelli}. This means that it has been verified that the 
electroweak gauge symmetry is indeed unbroken in the vertices of 
the theory, or in other words, the currents are indeed conserved 
\cite{altarelli}. 
   
It is well known that the Higgs mechanism is not the unique 
approach capable of generating the masses of elementary particles. In 
the following we mention some mechanisms of mass generation known in 
literature. One of these mechanisms is the dynamical breaking of the 
symmetry through radiative corrections at zero temperature developed 
by Coleman and Weinberg \cite{coleman}. Electroweak dynamical symmetry 
breaking, also called technicolor, provides an attractive mechanism 
for generating the gauge boson masses \cite{farhi}, while extended 
technicolor, walking technicolor, top condensation, and top-color 
assisted technicolor are mechanisms that have also been investigated 
\cite{farhi1}. Another known mechanism responsible for the ocurrence 
of mass terms, embedded in quantum field theory itself, is originated 
in the boundary conditions of fields and related to the topology of 
the manifold \cite{nogueira}. Dynamical mass generation at finite 
temperature using thermal corrections to the self energy is another 
alternative \cite{babu}-\cite{romeo}. However all these approaches 
seem to be unable to explain the spectrum of particle states as 
fully as the Higgs mechanism does.
 
In this paper we propose a new approach to generate the masses of 
elementary particles starting from the assumption that they 
correspond to the effective masses of the particles in an elementary 
quantum fluid at zero temperature. The physical source of mass 
generation in our approach is due to radiative corrections 
obtained in the framework of the quantum field theory at finite 
temperature and density. The elementary quantum fluid is 
constituted by massless quarks and massless leptons interacting 
through massless gluons, massless $W^{\pm}$ bosons, massless 
$Z^0$ bosons, massless photons and, possibly, massless scalar 
bosons. The fundamental effective model describing the 
structure and dynamics of this quantum fluid corresponds to the 
$SU(3)_C \times SU(2)_L \times U(1)_Y$ Standard Model without the 
Higgs Sector (SMWHS). A massless neutral scalar field can be 
included in the Lagrangian density of the SMWHS leading to Yukawa 
coupling terms between the massless fermions and the massless 
scalar boson with the same coupling constant ($Y$) for all the 
terms. This model describes the couplings of quark and leptons to 
the weak gauge bosons and the triple gauge boson vertices in 
accordance with the experimental observation that the electroweak 
currents are conserved \cite{altarelli}. In this model there are 
no Goldstone boson fields and the massless particle spectrum is 
symmetric under the gauge group. The elementary quantum fluid is 
characterized by non-vanishing fermionic chemical potentials 
($\mu_{fi}$) associated to the different quark and lepton flavours. 
The values of the different $\mu_{fi}$ are free parameters in 
this quantum fluid.  

We first calculate the fermionic effective masses starting from 
the one-loop self-energy at finite temperature ($T$) and 
non-vanishing fermionic chemical potentials ($\mu_f \neq 0$) and, 
in the rest frame, we identify these effective masses in the zero 
temperature limit with the fermion masses. Specifically, we obtain 
the fermion masses depending on the $\mu_{f_i}$ values and the gauge 
coupling constants of the fundamental effective model. Afterwards, 
starting from the known experimental values of the fermion masses, 
we obtain values for each $\mu_{fi}$ and from these ones we 
calculate, as an evidence of the consistency of our approach, the 
correct values for the $W^{\pm}$ and $Z^0$ gauge bosons masses 
using an analogous procedure as the one performed for the 
fermions. Specifically the bosonic effective masses are calculated 
starting from the one-loop bosonic polarization tensor at 
$T \neq 0$ and $\mu_f \neq 0$. The mixing between the $SU(2)_L$ 
gauge boson $(W_{\mu}^3)$ and the $U(1)_Y$ gauge boson $(B_{\mu})$ 
prevents the photon from acquiring mass. For the case where a 
massless neutral scalar field has been included in the Lagrangian 
density of the SMWHS, we also generate dynamically the mass of the 
neutral scalar boson starting from the scalar polarization tensor. 

In Section 2 we mention some relevant theoretical aspects which 
support our proposal. Before considering the real case of SMWHS 
we first calculate, in Section 3, the effective masses of the 
particles of an elementary quantum fluid at zero temperature that 
is described in structure and dynamics by a non-abelian gauge 
theory with Yukawa coupling terms and, we identify these 
effective masses with the rest masses of the particles in the 
vacuum. The effective masses are obtained at one-loop order and 
they are gauge invariant. In Section 4, using the SMWHS and 
following the same procedure as in Section 3, we dynamically 
generate the masses of the elementary fermions (quarks and 
leptons) and the electroweak bosons ($W^{\pm}$ and $Z^0$). These 
rest masses are obtained as functions of the $\mu_{f_i}$ values and 
the gauge coupling constants of the SMWHS. We use the experimental 
mass values for the fermions to obtain the $\mu_{f_i}$ values and 
we then calculate the $W^{\pm}$ and $Z^0$ gauge boson masses. The 
obtained gauge boson masses are in agreement with their experimental 
values. In Section 5, we add to the SMWHS Lagrangian density a 
massless neutral scalar field, including Yukawa coupling terms, and 
we then compute the $W^{\pm}$ and $Z^0$ gauge boson masses using an 
analogous manner as it was done in Section 4. Varying the top quark 
mass inside the experimental range imposes a range of possible 
values for the scalar boson mass. Our conclusions are summarized 
in Section 6.   

%%%%%%%%%%%%%%%%%%%%%%%%%%%%%%%%%%%%%%%%%%%%%%%%%%%%%%%%%%%%%%%%%%%%%%%

\seCtion{A theoretical overwiew}

To understand why our approach is a plausible way to generate 
consistently the masses of elementary particles we first mention 
some relevant theoretical aspects which support it. 

It has been suggested that the Universe and its ground state - the 
physical vacuum - may behave like a condensed matter system with a 
complicated and possibly degenerate ground-state 
\cite{hu}-\cite{jackiw}. This vacuum might be a richly structured 
medium, and an early indication of this feature could be the Dirac's 
sea \cite{wilczek}. As the physical vacuum is a complicated structure 
governed by locality and symmetry, one can learn how to analyse it by 
studying other systems with analogous properties like those found in 
condensed matter physics \cite{wilczek}. 

There are several phenomena in nature which have their origin in the 
properties of physical vacuum; one of them is the Casimir effect 
which describes an attractive force between two conducting plates 
\cite{casimir}. In quantum field theory the vacuum is a well-defined 
quantum state, specifically, the ground state of a system of fields. 
In this framework the Casimir effect is due to quantum fluctuations 
of the electromagnetic zero-point field in the intervening space. The 
same type of Casimir effect arises in condensed matter physics, 
particularly in correlated fluids, due to thermal and/or quantum 
fluctuations \cite{golestanian}. In this case the thermal fluctuations 
are modified by boundaries (membranes) resulting in finite-size 
corrections. Both cases can be studied for boundaries of arbitrary 
shape using the path integral formalism \cite{golestanian}. With the 
inclusion of quantum fluctuations the electromagnetic vacuum behaves, 
essentially, as a complex quantum fluid and modifies the motion of 
objects through it. In particular, the effective mass of a plate 
depends on its shape and becomes anisotropic \cite{golestanian}. When 
considering the analogy of the Casimir effect in condensed matter the 
following correspondence must be taken into account: the ground state 
of a quantum fluid corresponds to the vacuum of a quantum field theory 
\cite{golestanian}. This ground state is the quantum state of the fluid 
at zero temperature. In some cases the analogy between the quantum 
vacuum and the quantum fluid becomes exact. For instance, the 
low-energy fermionic and bosonic collective modes can correspond to the 
chiral fermions and gauge fields in quantum field theory \cite{volovik4}. 

The advantage of the quantum fluid is that the structure of the ground 
state is known, at least, in principle. For quantum fluids it is 
possible to calculate the phenomenological relevant parameters 
starting from a first principle microscopic theory. This effective 
theory, called the Theory of Everything, is ``a set of equations 
capable of describing all phenomena that have been observed" in these 
quantum systems \cite{laughlin}. For instance, in the context of this 
theory, Volovik has reported the calculation of a set of 
phenomenological parameters for the liquid $^4He$ at zero external 
pressure, obtaining a good agreement with experimental values 
\cite{volovik5}.

A phenomenological fact observed in the physical vacuum is that the 
elementary fermions (quarks and leptons) and the electroweak bosons 
($W^{\pm}$ and $Z^0$) are massive in the rest frame. We can think 
that the rest masses of these elementary particles reflect the 
physical properties of the vacuum. Our proposal is based on the two 
following ideas: (i) the ground state of a quantum fluid at zero 
temperature corresponds to the vacuum of quantum field theory 
\cite{golestanian}, (ii) it is possible to reproduce some observed 
properties of the physical vacuum, for instance the rest masses of 
elementary particles, starting from a first principle microscopic 
theory that describes an elementary quantum fluid \cite{laughlin}, 
for instance the MSWHS describing the fundamental interactions of 
quarks and leptons. Consequently, we can assume that the effective 
masses of the particles in the elementary quantum fluid at zero 
temperature correspond to the rest masses of the elementary 
particles in the physical vacuum.

%%%%%%%%%%%%%%%%%%%%%%%%%%%%%%%%%%%%%%%%%%%%%%%%%%%%%%%%%%%%%%%%%%%%%%%

\seCtion{Dynamical mass generation in a non-abelian \\ gauge theory}

\hspace{3.0mm}
In this section we calculate the effective masses of particles in an 
elementary quantum fluid at zero temperature described by a non-abelian 
gauge theory. This calculation is perform at one loop-order in the 
framework of the real time formalism of the quantum field theory at 
finite temperature and non-vanishing chemical potential. The quantum 
fluid is characterized by non-vanishing fermionic chemical potentials 
$\mu_{f_i} \not = 0$ where $f_i$ represents the different fermion 
species in the fluid. In this section we will take 
$\mu_{f_1}= \mu_{f_2}= \ldots = \mu_f$. The elementary quantum fluid 
is constituted by massless fermions interacting through massless gauge 
bosons and massless scalar bosons. The fundamental effective theory 
describing the structure and dynamics of this quantum fluid is the 
non-abelian gauge theory with Lagrangian density given by \cite{weldon}:

\be
{\cal L}=-\frac{1}{4} F_{A}^{\mu \nu} F_{\mu \nu}^{A} + \bar{\psi}_m 
\gamma^{\mu} \left( \delta_{mn} i \partial_{\mu} + 
g L_{mn}^{A} A_{\mu}^{A} \right) \psi_n + \frac{1}{2} \mbox{D}^{\mu} 
\phi \mbox{D}_{\mu} \phi + Y \bar{\psi}_m^L \Gamma_{mn}^i \psi_n^R 
\phi + H.C. , \la{lag}
\ee
where $A$ runs over the generators of the group and $m,n$ over the 
states of the fermion representation. The covariant derivative 
$(\mbox{D}_\mu)$ is $\mbox{D}_\mu = \delta_\mu + igT_A A_{\mu}^A$, 
being $T_A$ the generators of the $SU(N)$ gauge group and $g$ the 
gauge coupling constant. The last term of $(\ref{lag})$ represents 
the Yukawa interaction between the fermion fields and the scalar 
field $\phi$. The representation matrices $L_{mn}^{A}$ are 
normalized by $Tr(L^A L^B) = T(R) \delta^{AB}$ where $T(R)$ is 
the index of the representation. In the calculation of the 
fermionic self-energy appears $(L^A L^A)_{mn} = C(R) \delta_{mn}$, 
where $C(R)$ is the quadratic Casimir invariant of the 
representation \cite{weldon}.  

At finite temperature and density, the Feynman rules for vertices are
the same as those at $T=0$ and $\mu_f=0$, while the propagators in the
Feynman gauge for massless gauge bosons $D_{\mu \nu}(p)$, massless
scalars $D(p)$ and massless fermions $S(p)$ are \cite{kobes}:
\bea
D_{\mu \nu}(p) &=& -g_{\mu \nu} \left[ \frac{1}{p^2+i\epsilon} -i
\Gamma_b(p) \right],  \la{bp} \\
D(p) &=& \frac{1}{p^2+i\epsilon}-i{\Gamma}_b(p),  \la{ep} \\
S(p) &=& \frac{p{\hspace{-1.9mm}\slash}}{p^2+i \epsilon}+
i p{\hspace{-1.9mm} \slash}{\Gamma}_f(p),  \la{fp}
\eea
where $p$ is the particle four-momentum and the plasma temperature $T$
is introduced through the functions $\Gamma_b(p)$ and $\Gamma_f(p)$,
which are given by
\bea
\Gamma_b (p)= 2\pi \delta(p^2)n_b (p),  \la{db} \\
\Gamma_f (p)= 2\pi \delta(p^2)n_f (p),  \la{df}
\eea
with 
\bea
n_b (p) &=& \frac{1}{e^{(p\cdot u)/T}-1}, \la{nb}\\
n_f(p) &=& \theta(p\cdot u)n_{f}^{-}(p)+\theta(-p\cdot u)n_{f}^{+}(p),
\la{nf}
\eea
$n_b(p)$ being the Bose-Einstein distribution function. The Fermi-Dirac
distribution functions for fermions $n_{f}^{-}(p)$ and for anti-fermions
$n_{f}^{+}(p)$ are:
\bea
n_{f}^{\mp}(p)= \frac{1}{e^{(p\cdot u \mp \mu_f)/T}+1}.
\eea
In the distribution functions $(\ref{nb})$ and
$(\ref{nf})$, $u^{\alpha}$ is the four-velocity of the center-of-mass
frame of the dense plasma, with $u^\alpha u_\alpha =1$. 

\subsection{Fermionic self-energy and fermion mass}

\hspace{3.0mm}

For a non-abelian gauge theory with parity and chirality conservation 
the real part of the self-energy for a massless fermion is written as:
\be
\mbox{Re}\,\Sigma^{\prime}(K)=- aK{\hspace{-3.1mm}\slash}-b
u{\hspace{-2.1mm} \slash},  \la{tse}
\ee
$a$ and $b$ being the Lorentz-invariant functions and $K^{\alpha}$ the
fermion momentum. These functions depend on the Lorentz scalars
$\omega$ and $k$ defined by {\hspace {0.1 cm}} $\omega\equiv(K\cdot u)$
and $k\equiv[(K\cdot u)^2-K^2]^{1/2}$. Taking by convenience
$u^\alpha=(1,0,0,0)$ we have $K^2 =\omega^2-k^2$ and then, $\omega$ and 
$k$ can be interpreted as the energy and three-momentum, respectively.
Beginning with $(\ref{tse})$ it is possible to write:
\bea
a(\omega,k) &=& \frac{1}{4k^2} \left[ Tr(K{\hspace{-3.1mm}\slash}\,
\mbox{Re}\,\Sigma^{\prime})- \omega Tr(u{\hspace{-2.1mm}
\slash}\,\mbox{Re}\,\Sigma^{\prime}) \right],  \la{lifa} \\
b(\omega,k) &=& \frac{1}{4k^2} \left[ (\omega^2-k^2)
Tr(u{\hspace{-2.1mm}\slash}\,\mbox{Re}\,\Sigma^{\prime})- 
\omega Tr(K{\hspace{-3.1mm}\slash}\,\mbox{Re}\,\Sigma^{\prime}) 
\right].  \la{lifb}
\eea

The full fermion propagator, including only mass corrections, is given
by \cite{weldon1} 
\be
S(p)=\frac 1{K{\hspace{-3.1mm}\slash}-\mbox{Re}\,\Sigma ^{\prime }(K)}=
\frac{1}{r}\frac{\gamma^0 \omega n - \gamma_i k^i}{n^2 \omega^2 - k^2}, 
\la{pft}
\ee
where $n = 1 + b(\omega,k)/r\omega$ and $ r = 1 + a(\omega,k)$. The 
propagator poles can be found when: 
\be
\left[ 1 + \frac{b(w,k)}{1 + a(w,k)} \right]^2 w^2 - k^2 = 0.  \la{fdr0}
\ee
We observe in $(\ref{fdr0})$ that $n$ plays a role like the index of 
refraction in optics. To solve the equation $(\ref{fdr0})$, first it is 
required to calculate $a(\omega,k)$ and $b(\omega,k)$. These functions 
can be calculated from the relations $(\ref{lifa})$ and$(\ref{lifb})$ 
in terms of the real part of the fermionic self-energy. The one-loop 
diagrams that contribute to the fermionic self-energy are shown in 
Fig.$(1)$. The contribution to the fermionic self-energy from the gauge 
boson diagram shown in Fig.(1a) is given by
\be
\Sigma (K)=ig^2C(R)\int \frac{d^4p}{(2\pi )^4}D_{\mu \nu }(p)
{\gamma }^\mu S(p+K){\gamma }^\nu ,  \la{fse}
\ee
where $g$ is the interaction coupling constant and $C(R)$ is the
quadratic Casimir invariant of the representation.

Substituting $(\ref{bp})$ and $(\ref{fp})$ into $(\ref{fse})$, the
fermionic self-energy can be written as
$\Sigma(K)=\Sigma(0)+\Sigma^{\prime}(K)$, where $\Sigma(0)$ is the
zero-temperature and zero-density contribution and $\Sigma^{\prime}(K)$
is the finite-temperature and chemical potential contribution. It is
easy to see that: 
\bea
\Sigma(0)=-ig^2C(R) \int \frac{d^4 p}{(2\pi)^4} \frac{g_{\mu \nu}}{p^2}
\gamma^{\mu} \frac{p{\hspace{-1.9mm}\slash}+ K{\hspace{-3.1mm}\slash}}
{(p+K)^2} \gamma^{\nu}
\eea
and 
\bea
\Sigma^{\prime}(K)=2g^2 C(R) \int \frac{d^4 p}{(2\pi)^4}
(p{\hspace{-1.9mm}\slash}+ K{\hspace{-3.1mm}\slash}) \left[
\frac{\Gamma_b(p)}{(p+K)^2}-\frac{\Gamma_f(p+K)}{p^2}+i \Gamma_b(p)
\Gamma_f(p) \right].
\eea
Keeping only the real part $(\mbox{Re}\,\Sigma^{\prime}(K))$ of the
temperature and chemical potential contribution, we obtain:
\be
\mbox{Re}\,\Sigma^{\prime}(K)=2g^2C(R) \int \frac{d^4 p}{(2\pi)^4}
\left[(p{\hspace{-1.9mm}\slash}+ K{\hspace{-3.1mm}\slash}) \Gamma_b(p)+
p{\hspace{-1.9mm}\slash} \Gamma_f(p) \right] \frac{1}{(p+K)^2}.
\la{rse}
\ee
If we multiply $(\ref{rse})$ by either $K{\hspace{-3.1mm}\slash}$ or
$u{\hspace{-2.1mm}\slash}$, take the trace and perform the integrations
over $p_0$ and the two angular variables, the functions $(\ref{lifa})$ 
and $(\ref{lifb})$ can be written in the notation given in 
\cite{quimbay} as:
\bea
a(\omega,k)=g^2C(R)A(w,k,\mu_f), \la{aF0} \\
b(\omega,k)=g^2C(R)B(w,k,\mu_f), \la{bF0}
\eea
where the integrals over the modulus of the three-momentum 
$p= \vert \vec{p} \vert$, $A(\omega,k,\mu_f)$ and $B(\omega,k,\mu_f)$, 
are:

\bea
A(\omega,k,\mu_f) &=& \frac{1}{k^2}\int^\infty_0\frac{dp}{8\pi^2}
\left[ 2p-\frac{\omega p}{k}Log \left( \frac{\omega+k}{\omega-k} 
\right) \right] \left[2n_b(p)+n_f^-(p)+n_f^+(p) \right], \la{Alead} 
\nn \\
\\
B(\omega,k,\mu_f) &=& \frac{1}{k^2}  \nn \\
&\times& \int^\infty_0\frac{dp}{8\pi^2} \left[ \frac{p(\omega^2-k^2)}{k} 
Log \left( \frac{\omega+k}{\omega-k} \right) -2\omega p \right]
\left[2n_b(p)+n_f^-(p)+n_f^+(p)\right]. \la{Blead} \nn \\
\eea
The integrals $(\ref{Alead})$ and $(\ref{Blead})$ have been obtained 
in the high density limit, ${\it i. e.}$ $\mu_f >> k$ and 
$\mu_f >> \omega$, and keeping the leading terms in temperature and 
chemical potential \cite{morales}. Evaluating these integrals we 
obtain that $a(\omega,k)$ and $b(\omega,k)$ are given by:
\bea
a(\omega,k) &=& \frac{M_F^2}{k^2} \left[ 1-\frac{\omega}{2k}Log
\frac{\omega+k}{\omega-k}\right], \la{a1} \\
b(\omega,k) &=& \frac{M_F^2}{k^2} \left[ \frac{\omega^2-k^2}{2k}Log
\frac{\omega+k}{\omega-k}-\omega \right], \la{b1}
\eea
where the fermionic effective mass $M_F$ is:
\be
M_F^2(T, \mu_f)=\frac{g^2 C(R)}{8} \left( T^2+\frac{\mu_f^2}{\pi^2} 
\right).
\la{me}
\ee
The value of $M_F$ given by $(\ref{me})$ is in agreement with 
\cite{kajantie}-\cite{lebellac}. We are interested in the effective 
mass at $T=0$. For this case:
\be
M_F^2 (0, \mu_F)= M_{F_{\mu}}^2=\frac{g^2 C(R)}{8} 
\frac{\mu_f^2}{\pi^2}. \la{mef}
\ee 
Substituting $(\ref{a1})$ and $(\ref{b1})$ into $(\ref{fdr0})$, we 
obtain for the limit $k<<M_{F_{\mu}}$ that:
\be
\omega^2(k) = M_{F_{\mu}}^2 \left[ 1 + \frac{2}{3} 
\frac{k}{M_{F_{\mu}}} + \frac{5}{9} \frac{k^2}{M_{F_{\mu}}^2} + \dots 
\right] \la{dr1}
\ee
Owing to we have made the calculation at leading order in temperature 
and chemical potential, the dispersion relation is gauge independent 
\cite{morales}. It is very well known that the relativistic energy in 
the vacuum for a massive fermion at rest is $\omega^2 (0)= m_f^2$. At 
zero temperature the quantum fluid is in the ground state and it 
corresponds to the vacuum of quantum field theory \cite{golestanian}. 
It is clear from $(\ref{dr1})$ that if $k=0$ then 
$\omega^2 (0) = M_{F_{\mu}}^2$ and it is possible to identify the 
fermionic effective mass at zero temperature as the rest mass of the 
fermion. 

We note that the contribution to the real part of the fermionic 
self-energy from the generic scalar boson diagram shown in Fig.$(1b)$ 
has the same form as the gauge boson contribution $(\ref{rse})$. The 
factor $2g^2C(R)$ is replaced by $Y^2 C'$, where $Y$ is the 
Yukawa coupling constant and $C'$ is given in terms of the 
matrices of Clebsch-Gordan coefficients $\Gamma^i$ \cite{weldon}. 
The scalar boson contribution is proportional to 
$(\Gamma^i{\Gamma^{i}}^\dagger)_{mm'}\equiv C' \delta_{mm'}$.

Considering all contributions to the fermionic self-energy, we 
obtain that the fermion mass generated dynamically is:
\be
m_f^2 = \frac{(2g^2 C(R) + Y^2 C')}{16} \frac{\mu_f^2}{\pi^2}. 
\la{mas}
\ee
The fermion mass has been generated through the computation of 
the fermionic self-energy at one-loop order. The expression 
$(\ref{mas})$ is gauge invariant and in this expression both 
$\mu_f$ and $Y$ are free parameters.

\subsection{Bosonic polarization tensor and gauge boson mass}

\hspace{3.0mm}
The most general form of the polarization tensor which preserves 
the invariance under rotations, translations and gauge 
transformations is \cite{weldon2}:
\be
\Pi_{\mu \nu}(K) = P_{\mu \nu} \Pi_T (K) + Q_{\mu \nu} \Pi_L (K),
\la{pot}
\ee
where the Lorentz-invariant functions $\Pi_L$ and $\Pi_T$, which 
characterize the longitudinal and transverse modes respectively, 
are obtained by contraction:
\bea
\Pi_L(K)=-\frac{K^2}{k^2}u^{\mu}u^{\nu}\Pi_{\mu \nu}, \la{potl} \\
\Pi_T(K)=-\frac{1}{2}\Pi_L + \frac{1}{2}g^{\mu \nu} \Pi_{\mu \nu}. 
\la{pott}
\eea

The bosonic dispersion relations are obtained by looking at the 
poles of the full propagator wich is obtained by summing all the 
vacuum polarization insertions. The full bosonic propagator is 
\cite{weldon2}:
\be
D_{\mu \nu}(K) = \frac{Q_{\mu \nu}}{K^2 - \Pi_L (K)} + 
\frac{P_{\mu \nu}}{K^2 - \Pi_T (K)} - (\xi - 1)
\frac{K_\mu K_\nu}{K^4} , \la{fbp}
\ee
where $\xi$ is a gauge parameter. The gauge invariant dispersion 
relations, describing the two propagation modes, are found for:
\bea
K^2 - \Pi_L(K) = 0, \la{bdrl} \\
K^2 - \Pi_T(K) = 0. \la{bdrt}
\eea

The one-loop fermion contribution to the vacuum polarization from 
the diagram shown in Fig.(2a) is given by
\be
\Pi_{\mu \nu} (K)= i g^2 C(R) N_f \int \frac{d^4p}{(2\pi )^4} Tr 
\left[ {\gamma }_\mu S(p) {\gamma }_\nu S(p+K) \right],  \la{ptdf}
\ee
where $S$ are the propagators $(\ref{fp})$ and $N_f$ is the number 
of fermions in the fundamental representation. Substituting 
$(\ref{fp})$ into $(\ref{ptdf})$ the polarization tensor can be 
written as $\Pi_{\mu \nu}(K)= \Pi_{\mu \nu}(0)+
\Pi'_{\mu \nu}(K)$, where $\Pi_{\mu \nu}(0)$ is the 
zero-temperature and zero-density contribution and 
$\Pi'_{\mu \nu}(K)$ is the finite-temperature and chemical 
potential contribution.

It is easy to see that the real part of the finite-temperature and 
chemical potential contribution $\mbox{Re}\,\Pi'_{\mu \nu}(K)$ 
is given by
\be
\mbox{Re}\,\Pi'_{\mu \nu}(K) = \frac{g^2 C(R) N_f}{2} \int 
\frac{d^4p}{\pi^4} \frac{(p^2 + p \cdot K) g^{\mu \nu} - 2p^{\mu}
p^{\nu} - p^{\mu}K^{\nu} - p^{\nu}K^{\mu}}{(p+K)^2} \Gamma_f(p).
\la{rppt}
\ee
Substituting $(\ref{rppt})$ in $(\ref{potl})$ and $(\ref{pott})$ 
and keeping the leading terms in temperature and chemical potential, 
we obtain for the high density limit ($\mu_f >> k$ and 
$\mu_f >> \omega$) that:
\bea
\mbox{Re}\,\Pi'_L(K)=3 M_B^2 \left[ 1 - \frac{\omega}{2k} Log 
\frac{\omega+k}{\omega-k} \right], \la{rplo} \\
\mbox{Re}\,\Pi'_T(K)=\frac{3}{2} M_B^2 \left[ \frac{\omega^2}{k^2} + 
\left( 1 - \frac{\omega^2}{k^2} \right) \frac{\omega}{2k} Log 
\frac{\omega+k}{\omega-k} \right], \la{rptr}
\eea
where the bosonic effective mass $M_B$ is:
\be
M_B^2(T, \mu_f)= \frac{1}{6} N g^2 T^2 + \frac{1}{2} g^2 C(R) N_f 
\left[ \frac{T^2}{6} + \frac{\mu_f^2}{2 \pi^2} \right], \la{effb}
\ee
$N$ being the gauge group dimension. The effective mass $(\ref{effb})$ 
is in agreement with \cite{lebellac, braaten}. The bosonic effective 
mass at $T=0$ is:
\be
M_B^2(0, \mu_f)= M_{B_\mu}^2 =  \frac{g^2 C(R) N_f}{2} 
\frac{\mu_f^2}{\pi^2}, \la{bemT0}
\ee       
in agreement with the result obtained at finite density and $T=0$ 
theory \cite{altherr}. Substituting $(\ref{rplo})$ and $(\ref{rptr})$ 
into $(\ref{bdrl})$ and $(\ref{bdrt})$ for the limit $k<<M_{B_\mu}$, 
we obtain for the two propagation modes that:
\bea
\omega_L^2=M_{B_\mu}^2 + \frac{3}{5}k_L^2 \la{disl} \\
\omega_T^2=M_{B_\mu}^2 + \frac{6}{5}k_T^2. \la{dist} 
\eea    
We note that $(\ref{disl})$ and $(\ref{dist})$ have the same value 
when the three-momentum goes to zero. It is clear from $(\ref{disl})$ 
and $(\ref{dist})$ that for $k=0$ then $\omega^2(0)=M_{B_\mu}^2$ and 
it is possible to recognize the bosonic effective mass at zero 
temperature as the rest mass of the gauge boson:
\be
m_b^2 = M_{B_\mu}^2 = \frac{g^2 C(R) N_f}{2} \frac{\mu_f^2}{\pi^2}. 
\la{bosmas}
\ee

\subsection{Scalar polarization tensor and scalar boson mass}

\hspace{3.0mm}
The full scalar boson propagator is:
\be
D(K) = \frac{1}{K^2 - \Pi(K)} \la{sbp}
\ee
where the dispersion relations are obtained for:
\be
K^2 - \Pi (K) = 0, \la{sbdr} 
\ee

The one-loop fermion contribution to the scalar polarization tensor 
from the diagram shown in Fig.(2b) is given by
\be
\Pi (K)= i Y^2 N_f \int \frac{d^4p}{(2\pi )^4} Tr 
\left[ S(p) S(p+K) \right],  \la{sbpt}
\ee
where $S$ are the propagators $(\ref{fp})$ and $N_f$ is the number 
of fermions in the fundamental representation. Substituting 
$(\ref{fp})$ into $(\ref{sbpt})$, the polarization tensor can be 
written as $\Pi (K)= \Pi (0)+ \Pi' (K)$, where $\Pi (0)$ is the 
zero-temperature and zero-density contribution and $\Pi' (K)$ is 
the finite-temperature and chemical potential contribution.

The real part of the finite-temperature and chemical potential 
contribution $\mbox{Re}\,\Pi' (K)$ 
is given by
\be
\mbox{Re}\,\Pi' (K) = \frac{Y^2 N_f}{2} \int \frac{d^4p}{\pi^4} 
\frac{(p^2 + p \cdot K)}{(p+K)^2} \Gamma_f (p). \la{rsbpt}
\ee

Substituting $(\ref{df})$ in $(\ref{rsbpt})$ and keeping the 
leading terms in temperature and chemical potential, we obtain 
for the high density limit ($\mu_f >> k$ and $\mu_f >> \omega$) 
that:
\be
\mbox{Re}\,\Pi'(K)= Y^2 N_f \left[ \frac{T^2}{6} + 
\frac{\mu_f^2}{2 \pi^2} \right] = M_S^2(T, \mu_f), \la{sem}
\ee
$M_S$ being the scalar effective mass. For $T=0$, the scalar 
effective mass is:
\be
M_S^2(0, \mu_f)= M_{S_\mu}^2 =  \frac{Y^2 N_f}{2} 
\frac{\mu_f^2}{\pi^2}. \la{sem0}
\ee       
The dispersion relation at $T= 0$ is obtained from $(\ref{sbdr})$ 
being:
\be
w^2= k^2 + M_{S_\mu}^2  \la{drsb}
\ee    
If we put $k=0$ in $(\ref{drsb})$ then $w^2(0)=M_{S_\mu}^2$ and 
it is possible to identify the scalar effective mass at zero 
temperature as the rest mass of the scalar boson:
\be
m_s^2 = M_{S_\mu}^2 = \frac{Y^2 N_f}{2} \frac{\mu_f^2}{\pi^2}. 
\la{scamas}
\ee

%%%%%%%%%%%%%%%%%%%%%%%%%%%%%%%%%%%%%%%%%%%%%%%%%%%%%%%%%%%%%%%%%%%%

\seCtion{Dynamical mass generation in the SMWHS}

\hspace{3.0mm}
In this section, following the same procedure as in Section 3, we 
generate dynamically the masses of the elementary particles of the 
SMWHS. To do it, we first calculate the effective masses of the 
particles in an elementary quantum fluid described in structure and 
dynamics by the SMWHS. The elementary quantum fluid is constituted 
by massless quarks and massless leptons interacting through 
massless gluons, massless $W^{\pm}$ bosons, massless $Z^0$ bosons 
and massless photons. The quantum fluid is characterized by 
non-vanishing chemical potentials associated to the different 
fermion flavour species. We consider for quarks
$\mu_u \not = \mu_d \not =\mu_c \not =... \not =0$, for charged
leptons $\mu_e \not = \mu_{\mu} \not =\mu_{\tau} \not =0$ and for
neutrinos $\mu_{\nu_e} \not = \mu_{\nu_{\mu}} \not =\mu_{\nu_{\tau}}
\not =0$. The $SU(2)_L \times U(1)_Y$ Standard Model without Higgs 
Sector is described by the Lagrangian density:
\be
{\cal L}_{eff} = {\cal L}_{YM} + {\cal L}_{FB} + {\cal L}_{GF} + 
{\cal L}_{FP}, \la{ldm}  
\ee
where ${\cal L}_{YM}$ is the Yang-Mills Lagrangian density, 
${\cal L}_{FB}$ is the fermionic-bosonic Lagrangian density, 
${\cal L}_{GF}$ is the gauge fixing Lagrangian density and 
${\cal L}_{FP}$ is the Fadeev-Popov Lagrangian density. The 
${\cal L}_{YM}$ is given by
\be
{\cal L}_{YM} = -\frac{1}{4} W_{A}^{\mu \nu} W_{\mu \nu}^{A} 
-\frac{1}{4} F^{\mu \nu} F_{\mu \nu} \la{lym}
\ee
where $W_{\mu \nu}^{A} = \partial_{\mu} W_{\nu}^A - 
\partial_{\mu} W_{\mu}^A + g_w F^{ABC} W_{\mu}^B W_{\nu}^C $ is 
the energy-momentum tensor associated to the $SU(2)_L$ group 
and $ F_{\mu \nu} = \partial_{\mu}B_{\nu} - \partial_{\mu} 
B_{\mu}$ is the one associated to the $U(1)_Y$ group. The 
${\cal L}_{FB}$ is written as:   
\be
{\cal L}_{FB} = i \bar{\mbox{L}} \gamma^{\mu}\mbox{D}_\mu 
\mbox{L} + i \psi^i_R \gamma^{\mu}\mbox{D}_\mu \psi^i_R 
+ i \psi^I_R \gamma^{\mu}\mbox{D}_\mu \psi^I_R, 
\la{lafb}  
\ee  
where $\mbox{D}_\mu \mbox{L} = ( \partial_{\mu} + ig' Y_L B_\mu/2 
+ ig T_i W_{\mu}^i ) \mbox{L}$ and $\mbox{D}_\mu \mbox{R} = 
(\partial_{\mu} + ig' Y_R B_\mu/2 ) \mbox{R}$, being $g_w$ the 
gauge coupling constant associated to the $SU(2)_L$ group, $g_e$ 
the one associated to the $U(1)_Y$ group, $Y_L = -1$, $Y_R = -2$ 
and $T_i = \sigma_i /2$. The $SU(2)_L$ left-handed doublet 
$(\mbox{L})$ is given by
\be
\mbox{L}= {\psi^i \choose {\psi^I}}_L .
\ee

For a non-abelian gauge theory with parity violation and quirality
conservation like the SMWHS, the real part of the self-energy for 
a massless fermion is:
\be
\mbox{Re}\,\Sigma'(K)=- K{\hspace{-3.1mm}\slash}(a_{L}P_L +a_{R}P_R)-
u{\hspace{-2.2mm}\slash}(b_{L}P_L +b_{R}P_R),
\ee
where $P_L \equiv\frac{1}{2}(1-\gamma_5)$ and
$P_R \equiv\frac{1}{2}(1+\gamma_5)$ are the left- and right-handed
chiral projectors respectively. The functions $a_L$, $a_R$, $b_L$
and $b_R$ are the chiral projections of the Lorentz-invariant
functions $a$, $b$ and they are defined in the following way:
\bea
a &=& a_L P_L + a_R P_R, \\
b &=& b_L P_L + b_R P_R.
\eea
The inverse fermion propagator is given by
\be
S^{-1}(K)= {\cal L}{\hspace{-2.5mm}\slash} P_L +
\Re{\hspace{-2.5mm}\slash} P_R \la{ifp}
\ee
where:
\bea
{\cal L}^{\mu} &=& ( 1 + a_L) K^{\mu} + b_L u^{\mu} \\
{\Re}^{\mu} &=& ( 1 + a_R) K^{\mu} + b_R u^{\mu}
\eea
The fermion propagator follows from the inversion of $(\ref{ifp})$:
\bea
S=\frac{1}{D}\left[\left({\cal L}^2\Re{\hspace{-2.5mm}
\slash}\right)P_L + \left(\Re^2{\cal L}{\hspace{-2.5mm}\slash}
\right)P_R \right].\la{p1}
\eea
being $D(\omega,k)={\cal L}^2 {\Re}^2$. The poles of the propagator
correspond to values $\omega$ and $k$ for which the determinat $D$ 
in (\ref{p1}) vanishes:
\be
{\cal L}^2 {\Re}^2 =0.\la{d}
\ee
In the rest frame of the dense plasma $u=(1,\vec 0)$, Eq.$(\ref{d})$
leads to the fermionic dispersion relations for a chirally invariant 
gauge theory with parity violation, as the case of the SMWHS . Thus, 
the fermionic dispersion relations for this case are given by
\bea
\left[ \omega (1+a_L)+b_L \right]^2- k^2 \left[ 1+a_L \right]^2 &=& 0,
\la{dra}\\
\left[ \omega (1+a_R)+b_R \right]^2-k^2 \left[ 1+a_R \right]^2 &=& 0.
\la{drb}
\eea
Left- and right-handed components of the fermionic dispersion relations 
obey decoupled relations. The Lorentz invariant functions $a(\omega,k)$ 
and $b(\omega,k)$ are calculated from the expressions $(\ref{lifa})$ 
and $(\ref{lifb})$ through the real part of the fermionic self-energy. 
This self-energy is obtained adding all the posibles gauge boson 
contributions admited by the Feynman rules of the SMWHS. 

We will work in the basis of gauge bosons given by $B_\mu$, $W_{\mu}^3$, 
$W_{\mu}^{\pm}$, where the charged electroweak gauge boson is 
$W_{\mu}^{\pm} = (W_{\mu}^1 \mp iW_{\mu}^2)/ \sqrt{2}$. The diagrams
with an exchange of $W^{\pm}$ gauge bosons induce a flavour change 
in the incoming fermion $i$ to a different outgoing fermion $j$.

For the quark sector, in the case of the flavour change contributions
mentioned, the flavor $i$ $(I)$ of the internal quark (inside the loop) 
runs over the up $(i)$ or down $(I)$ quarks flavours according to the 
type of the external quark (outside the loop). Owing to each 
contribution to the quark self-energy is proportional to 
$(\ref{Alead})$-$(\ref{Blead})$, the functions $a_L$, $a_R$, 
$b_L$ and $b_R$ are given by
\bea
a_L(\omega,k)_{ij} &=&
[f_{S}+f_{B}+f_{W^3}] A(\omega,k,\mu_{i}) + \sum_{I}
f_{W^{\pm}} A(\omega,k,\mu_{I}), \la{al} \\
b_L(\omega,k)_{ij} &=& 
[f_{S}+f_{B}+f_{W^3}] B(\omega,k,\mu_{i}) + \sum_{I}
f_{W^{\pm}} B(\omega,k,\mu_{I}), \la{bl} \\ 
a_R(\omega,k)_{ij} &=&
[f_{S}+f_{B}] A(\omega,k,\mu_{i}), \la{ar} \\
b_R(\omega,k)_{ij} &=& 
[f_{S}+f_{B}] B(\omega,k,\mu_{i}). \la{br}
\eea
The coefficients $f$ are:
\bea
f_{S} &=& \frac{4}{3}g_s^2 \delta_{ij}, \\
f_{B} &=& \frac{1}{4} g_w^2 \delta_{ij}, \\
f_{W^3}&=& \frac{1}{4} g_e^2 \delta_{ij}, \\
f_{W^{\pm}} &=& \frac{1}{2} g_w^2 K_{il}^{+}K_{lj},
\eea
where ${\it K}$ represents the CKM matrix and $g_s$ is the strong
coupling constant. The integrals $A(\omega,k,\mu_f)$ and 
$B(\omega,k,\mu_f)$ have been obtained in the high density limit 
$(\mu_f >> k$ and $\mu_f >> \omega )$ and keeping the leading 
terms in temperature and chemical potential. They are given by
\bea
A(\omega,k,\mu_f) &=& \frac{1}{8 k^2} \left( T^2+ \frac{\mu_f^2}
{\pi^2} \right)\left[1-\frac{\omega}{2k} Log \frac{\omega+k}
{\omega-k} \right] \la{ALR},
\\
B(\omega,k,\mu_f) &=& \frac{1}{8 k^2} \left( T^2+ \frac{\mu_f^2}
{\pi^2}\right)\left[\frac{\omega^2-k^2}{2k} Log \frac{\omega+k}
{\omega-k}-\omega \right].\la{BLR}
\eea

The chiral projections of the Lorentz-invariant functions are:
\bea
a_{L}(\omega,k)_{ij} &=& \frac{1}{8k^2}\left[1-F(\frac{\omega}{k})
\right]\left[l_{ij}(T^2+\frac{\mu_i^2}{\pi^2})+ c_{ij}(T^2+
\frac{\mu_i^2}{\pi^2})\right], \la{aL} \\
b_{L}(\omega,k)_{ij} &=& -\frac{1}{8k^2}\left[\frac{\omega}{k}+
(\frac{k}{\omega}-\frac{\omega}{k})F(\frac{\omega}{k})\right]\left
[l_{ij}(T^2+\frac{\mu_i^2}{\pi^2})+ c_{ij}(T^2+\frac{\mu_i^2}{\pi^2}) 
\right], \nn \la{bL} \\
\\
a_{R}(\omega,k)_{ij} &=& \frac{1}{8k^2}\left[1-F(\frac{\omega}{k})
\right]\left[r_{ij}(T^2+\frac{\mu_i^2}{\pi^2}) \right], \la{aR} \\
b_{R}(\omega,k)_{ij} &=& -\frac{1}{8k^2}\left[\frac{\omega}{k}+
(\frac{k}{\omega}-\frac{\omega}{k})F(\frac{\omega}{k})\right]
\left[r_{ij}(T^2+\frac{\mu_i^2}{\pi^2})\right], \la{bR}
\eea
where $F(x)$ is
\be
F(x)=\frac{x}{2} Log \left(\frac{x+1}{x-1}, \right)
\ee
and the coefficients are given by
\bea
l_{ij} &=& \left( \frac{4}{3}g_s^2 + \frac{1}{4}g_w^2 + 
\frac{1}{4}g_e^2 \right)\delta_{ij}, \\
c_{ij} &=& \sum_l \left(\frac{g_w^2}{2}\right) K_{il}^+ K_{lj}, \\
r_{ij} &=& \left(\frac{4}{3}g_s^2 + \frac{1}{4}g_e^2 \right)
\delta_{ij}.
\eea
Substituting $(\ref{aL})$-$(\ref{bL})$ into $(\ref{dra})$, and
$(\ref{aR})$-$(\ref{bR})$ into $(\ref{drb})$, we obtain for the 
limit $k << M_{(i,I)_{L,R}}$ that:
\be
\omega^2(k) = M_{(i,I)_{L,R}}^2 \left[ 1 + \frac{2}{3} 
\frac{k}{M_{(i,I)_{L,R}}} + \frac{5}{9} \frac{k^2}{M_{(i,I)_{L,R}}^2} 
+ \dots 
\right], \la{drsmlr}
\ee
where:
\bea
M_{(i,I)_L}^2(T,\mu_f) &=& (l_{ij}+c_{ij}) \frac{T^2}{8} + l_{ij} 
\frac{\mu_{(i,I)_L}^2}{8 \pi^2} + c_{ij} \frac{\mu_{(I,i)_L}^2}
{8 \pi^2}, \la{emL}\\
M_{(i,I)_R}^2(T,\mu_f)  &=& (r_{ij}+d_{ij}) \frac{T^2}{8} + r_{ij} 
\frac{\mu_{(i,I)_R}^2}{8 \pi^2}. \la{emR}
\eea
We are interested in the effective masses at $T=0$. For this case:
\bea
M_{(i,I)_L}^2(0,\mu_f) &=& l_{ij} \frac{\mu_{(i,I)_L}^2}{8 \pi^2} + 
c_{ij} \frac{\mu_{(I,i)_L}^2}{8 \pi^2}, \la{emLT0}\\
M_{(i,I)_R}^2(0,\mu_f)  &=& r_{ij} \frac{\mu_{(i,I)_R}^2}{8 \pi^2}. 
\la{emRT0}
\eea

The fermionic dispersion relation for the lepton sector are similar 
to the relations $(\ref{drsmlr})$, even though the effective masses 
changing. For this sector $g_s = 0$ and non-exist mixing between 
the charge lepton flavours. Then, for the lepton sector, the 
coefficients $l$, $c$ and $r$ in $(\ref{emLT0})$ and $(\ref{emRT0})$ 
are:
\bea
l &=& \left( \frac{1}{4}g_w^2 + \frac{1}{4}g_e^2 \right), \\
c &=& \left( \frac{1}{2} g_w^2 \right), \\
r &=& \left(\frac{1}{4}g_e^2 \right). 
\eea
For the neutrino sector the coefficient $r$ is zero since there are 
not right-handed neutrinos in the physical spectrum. 

Following the same argument as in Section 3, we can identify the 
effective masses of the particles in the elementary quantum fluid 
at zero temperature as the masses of the elementary particles at 
rest. Coming from the left-handed and right-handed representations, 
respectively, we find for the quark sector that the rest masses are:
\bea
m_{i}^2 = \left[ \frac{4}{3}g_s^2 + \frac{1}{4}g_w^2 + \frac{1}{4}
g_e^2 \right] \frac{\mu_{i_L}^2}{8 \pi^2} + \left[ \frac{1}{2}g_w^2 
\right] \frac{\mu_{I_L}^2}{8 \pi^2}, \la{uqpl} \\
m_{I}^2 = \left[ \frac{4}{3}g_s^2 + \frac{1}{4}g_w^2 + \frac{1}{4}
g_e^2 \right] \frac{\mu_{I_L}^2}{8 \pi^2} + \left[ \frac{1}{2}g_w^2 
\right] \frac{\mu_{i_L}^2}{8 \pi^2},  \la{dqpl}    
\eea
and
\bea
m_{i}^2 = \left[ \frac{4}{3}g_s^2 + \frac{1}{4} g_e^2 \right] 
\frac{\mu_{i_R}^2}{8 \pi^2}, \la{uqpr} \\
m_{I}^2 = \left[ \frac{4}{3}g_s^2 + \frac{1}{4} g_e^2 \right] 
\frac{\mu_{I_R}^2}{8 \pi^2}, \la{dqpr}    
\eea 
where the couple of index $(i,I)$ running over the quarks $(u, d)$, 
$(c, s)$ and $(t, b)$. 

For the lepton sector we find that the rest masses are:
\bea
m_{i}^2 = \left[ \frac{1}{4}g_w^2 + \frac{1}{4} g_e^2 \right] 
\frac{\mu_{i_L}^2}{8 \pi^2} + \left[ \frac{1}{2}g_w^2 
\right] \frac{\mu_{I_L}^2}{8 \pi^2}, \la{nlpl} \\
m_{I}^2 = \left[ \frac{1}{4}g_w^2 + \frac{1}{4} g_e^2 \right] 
\frac{\mu_{I_L}^2}{8 \pi^2} + \left[ \frac{1}{2}g_w^2 
\right] \frac{\mu_{i_L}^2}{8 \pi^2},  \la{elpl}    
\eea
and
\bea
m_{I}^2 = \left[ \frac{1}{4} g_e^2 \right] 
\frac{\mu_{I_R}^2}{8 \pi^2}, \la{elpr}    
\eea
where the couple of index $(i,I)$ running over the leptons 
$(\nu_e, e)$, $(\nu_{\mu}, \mu)$ and $(\nu_{\tau}, \tau)$.

We have generated dynamically the masses of the fermionic spectrum. 
These masses depended on the $\mu_{f_i}$ values and the gauge 
coupling constants of the SMWHS. Because the $\mu_{f_i}$ values 
are free parameters, we can compute these starting from the known 
experimental values for the fermion masses.

To evaluate the bosonic polarization tensor associated with the 
$W^{\pm}_\mu$, $W^3_\mu$, $B_\mu$ gauge boson propagators, we 
follow the same procedure as in Section 3. We find that the 
gauge boson masses generated dynamically are:
\bea
M_{W^\pm}^2 = \frac{g_w^2}{4} \sum_{f=1}^{12} \frac{ \mid \mu_{f_L}^2 
\mid}{2 \pi^2}, \la{mw+-} \\
M_{W^3}^2 = \frac{g_w^2}{4} \sum_{f=1}^{12} \frac{ \mid \mu_{f_L}^2 
\mid}{2 \pi^2}, \la{mw3} \\
M_{B}^2 = \frac{g_e^2}{4} \sum_{f=1}^{12} \frac{ \mid \mu_{f_L}^2 
\mid}{2 \pi^2}, \la{mb}
\eea
where the sum running over the twelve $\mu_{f_L}$ values associated 
with the elementary fermions of the physical spectrum. 
    
Starting from the expressions $(\ref{uqpl})$, $(\ref{dqpl})$, 
$(\ref{nlpl})$, $(\ref{elpl})$, using the experimental central 
mass values for the fermions \cite{groom} $m_u = 0.003$, $m_d = 0.006$, 
$m_c = 1.25$, $m_c = 0.122$, $m_b = 4.2$, $m_{\nu_e} = 0$, 
$m_e = 0.0005$, $m_{\nu_{\mu}} = 0$, $m_\mu = 0.1056$, 
$m_{\nu_{\tau}} = 0$, $m_\tau = 1.777$, $m_t = 174.3 \pm 5.1$ GeV, 
and $g_s = 1.22029$, $g_w = 0.642343$ , $g_e = 0.117906$, we 
obtain the $\mu_{f_L}$ values for all the fermion flavour species. 
Substituting these $\mu_{f_L}$ values in the expressions 
$(\ref{mw+-})$, $(\ref{mw3})$ and $(\ref{mb})$, we obtain 
\bea
M_{W^{\pm}}= M_{W^{3}}= 81.0613 \pm 2.3667 \,{\mbox GeV}, \la{masw} \\ 
M_{B}= 43.3326 \pm 1.2652 \,{\mbox GeV}, \la{masb}
\eea  
where the $M_W$ value computed is in agreement with the experimental 
value.

By physical reasons $W_{\mu}^3$ and $B_\mu$ gauge bosons are mixed. 
After diagonalization of mass matrix, we get the physical fields 
$A_\mu$ and $Z_\mu$ corresponding to the massless photon and the 
neutral $Z^0$ boson of mass $M_Z$ respectively, with the relations 
\cite{weinberg, pestieau}:
\bea 
M_Z^2 = M_W^2 + M_B^2 , \la{masz} \\
cos \theta_w = \frac{M_W}{M_Z} \hspace{3.0mm} ,  \hspace{3.0mm} 
sen \theta_w = \frac{M_B}{M_Z}, \la{mix}
\eea
where $\theta_w$ is the weak mixing angle:
\bea
Z_{\mu}^0 = B_\mu sen \theta_w - W_{\mu}^3 cos \theta_w , \la{zbw} \\
A_{\mu} = B_\mu cos \theta_w + W_{\mu}^3 sen \theta_w . \la{abw}
\eea     
Substituting $(\ref{masw})$ and $(\ref{masb})$ into $(\ref{masz})$
we obtain that:
\be
M_{Z}= 91.9166 \pm 2.6837 \,{\mbox GeV}, \la{masZ0} 
\ee   
in agreement with the experimental value. We observe that using a top 
mass value of $172.9159$ GeV in the expressions $(\ref{uqpl})$, 
$(\ref{dqpl})$, $(\ref{nlpl})$ and $(\ref{elpl})$, we obtain for 
the $W^{\pm}$ and $Z^0$ boson masses the experimental central values: 
$M_W = 80.419$ and $M_Z = 91.1882$ GeV.

%%%%%%%%%%%%%%%%%%%%%%%%%%%%%%%%%%%%%%%%%%%%%%%%%%%%%%%%%%%%%%%%%%%%%%

\seCtion{SMWHS plus a massless neutral scalar boson}

\hspace{3.0mm}
In this section, we add to SMWHS Lagrangian density a massless 
neutral scalar field with Yukawa coupling terms between the 
massless fermions and the massless scalar boson. The Yukawa coupling 
terms are taken equals for all the fermions, ${\it i. e.}$ the 
coupling constant (Y) is the same for all the couplings. Following 
the same procedure as in section 4, we generate dynamically the 
masses of the elementary particles. In this case the elementary 
quantum fluid is constituted by massless quarks and massless leptons 
interacting through massless gluons, massless $W^{\pm}$ bosons, 
massless $Z^0$ bosons, massless photons and massless scalar bosons. 
The quantum fluid is characterized by non-vanishing chemical 
potentials associated to the different fermion flavour species 
$\mu_{f_i} \neq 0$. The elementary effective model is described 
by the Lagrangian density:
\be
{\cal L}'_{eff} = {\cal L}_{eff} + {\cal L}_{SB}, \la{ldmps}  
\ee
where ${\cal L}_{eff}$ is the SMWHS Lagrangian density 
$(\ref{ldm})$ and ${\cal L}_{SB}$ is the Scalar Boson Lagrangian 
density given by
\be
{\cal L}_{SB} = \left( {\mbox D}^\mu \phi \right)^{\dagger}
\left( {\mbox D}_\mu \phi \right) + {\cal L}_{Y}, \la{lsby}
\ee
where $\mbox{D}_\mu = \partial_\mu - ig' B_\mu/2 + 
ig \sigma_i W_{\mu}^i/2$ and the Yukawa Lagrangian density 
${\cal L}_{Y}$ is:
\be
{\cal L}_{Y} = Y \left[ \bar \psi_R^I (\phi^{\dagger} \mbox{L}) + 
(\bar{\mbox{L}} \phi)\psi_R^I + \bar \psi_R^i (\phi^{\dagger}_c 
\mbox{L}) + (\bar{\mbox{L}} \phi_c)\psi_R^i \right], \la{layuca}
\ee
being $Y$ the Yukawa coupling constant. The doublets $\phi$ and 
$\phi^{\dagger}$ are given by
\bea
\phi= {\phi^+ \choose {\phi^0}}_L \la{doub}, \\
\phi^{\dagger}= i \sigma_2 \phi^* = {\phi^0 \choose {-\phi^-}}_L 
\la{doubco}. 
\eea
We can also introduce in the Lagrangian density $(\ref{lsby})$ 
a scalar potential leading to auto-interacting scalar terms, but 
these vertices will not affect the results that we obtain in this 
section. The Lagrangian density $(\ref{layuca})$ can be written 
as: 
\bea
{\cal L}_{Y} = Y \sum_{e, \nu, d, u}^3 \left[ \bar e_R \phi^0 e_L 
+ \bar {\nu_e}_R \phi^0 {\nu_e}_L + \bar e_L \phi^0 e_R + 
\bar {\nu_e}_L \phi^0 {\nu_e}_R  \right. \nn \\
\left. + \bar d_R \phi^0 d_L + \bar u_R \phi^0 u_L + \bar d_L 
\phi^0 d_R + \bar u_L \phi^0 u_R \right], \la{layu}
\eea   
where the sum running over the three families of fermions and we 
have chosen a particular doublet $\phi$ of the form:
\be
\phi= {0 \choose {\phi^0}}_L \la{doub} 
\la{doubco}. 
\ee
With this selection, we have take the charged scalar boson as zero.
Following the same procedure as in Section 4, we generate dynamically 
the masses of the elementary particles. Coming from the left-handed 
and right-handed representations, respectively, we find for the 
quark sector that the rest masses are:
\bea
m_{i}^2 = \left[ \frac{4}{3}g_s^2 + \frac{1}{4}g_w^2 + \frac{1}{4}
g_e^2 + \frac{Y^2}{2} \right] \frac{\mu_{i_L}^2}{8 \pi^2} + 
\left[ \frac{1}{2}g_w^2 \right] \frac{\mu_{I_L}^2}{8 \pi^2}, 
\la{uqplws} \\
m_{I}^2 = \left[ \frac{4}{3}g_s^2 + \frac{1}{4}g_w^2 + \frac{1}{4}
g_e^2 + \frac{Y^2}{2} \right] \frac{\mu_{I_L}^2}{8 \pi^2} + 
\left[ \frac{1}{2}g_w^2 \right] \frac{\mu_{i_L}^2}{8 \pi^2},  
\la{dqplws}    
\eea
and
\bea
m_{i}^2 = \left[ \frac{4}{3}g_s^2 + \frac{1}{4} g_e^2 + \frac{Y^2}{2} 
\right] \frac{\mu_{i_R}^2}{8 \pi^2}, \la{uqprws} \\
m_{I}^2 = \left[ \frac{4}{3}g_s^2 + \frac{1}{4} g_e^2 + \frac{Y^2}{2} 
\right] \frac{\mu_{I_R}^2}{8 \pi^2}, \la{dqprws}    
\eea 
where the couple of index $(i,I)$ running over the quarks $(u, d)$, 
$(c, s)$ and $(t, b)$. 

For the lepton sector we find that the rest masses are:
\bea
m_{i}^2 = \left[ \frac{1}{4}g_w^2 + \frac{1}{4} g_e^2 + \frac{Y^2}{2} 
\right] \frac{\mu_{i_L}^2}{8 \pi^2} + \left[ \frac{1}{2}g_w^2 
\right] \frac{\mu_{I_L}^2}{8 \pi^2}, \la{nlplws} \\
m_{I}^2 = \left[ \frac{1}{4}g_w^2 + \frac{1}{4} g_e^2 + \frac{Y^2}{2} 
\right] \frac{\mu_{I_L}^2}{8 \pi^2} + \left[ \frac{1}{2}g_w^2 
\right] \frac{\mu_{i_L}^2}{8 \pi^2},  \la{elplws}    
\eea
and
\bea
m_{I}^2 = \left[ \frac{1}{4} g_e^2 + \frac{Y^2}{2}\right] 
\frac{\mu_{I_R}^2}{8 \pi^2}, \la{elprws}    
\eea
where the couple of index $(i,I)$ running over the leptons 
$(\nu_e, e)$, $(\nu_{\mu}, \mu)$ and $(\nu_{\tau}, \tau)$.

The fermion masses depended on the $\mu_{f_i}$ values and the 
coupling constants. We can compute the $\mu_{f_i}$ values from 
the known experimental values of the fermion masses. Starting from 
the expressions $(\ref{uqplws})$, $(\ref{dqplws})$, 
$(\ref{nlplws})$, $(\ref{elplws})$, using the experimental central 
mass values for the fermions, substituting the obtained $\mu_{f_L}$ 
values in the expressions $(\ref{mw+-})$, $(\ref{mw3})$ and 
$(\ref{mb})$, and using $(\ref{masz})$, we find that the 
experimental mass central values for the $W^{\pm}$ and $Z^0$ 
gauge bosons can be obtained using a masss value for the top quark 
mass in the range $169.2$ GeV $< m_{t} < 178.6$ GeV and a Yukawa 
coupling constant in the range $0 < Y < 0.607$. This range for 
the top quark mass implies that the scalar boson mass must be in 
the range $0 < M_{H} < 152$ GeV as it is shown in the Fig.(3).

The same mechanism of mass dynamical generation acts on the scalar 
boson given a mass $M_H$ which can be calculated from:
\be
M_H^2 = Y^2 \sum_{f=1}^{12} \frac{ \mid \mu_{f_L}^2 
\mid}{2 \pi^2}, \la{mhsb} \la{mb}
\ee
where the sum running over the $\mu_{f_L}$ values that we have 
computed before.

%%%%%%%%%%%%%%%%%%%%%%%%%%%%%%%%%%%%%%%%%%%%%%%%%%%%%%%%%%%%%%%%%%%%%%

\seCtion{Conclusions}

\hspace{3.0mm}

We have introdused a new approach to generate dinamically the masses 
of elementary particles in the $SU(3)_C \times SU(2)_L \times U(1)_Y$ 
Standard Model without Higgs Sector (SMWHS). Our proposal have been 
founded on the two following ideas: (i) the ground state of a quantum 
fluid at zero temperature corresponds to the vacuum of quantum field 
theory \cite{golestanian}, (ii) it is possible to reproduce some 
observed properties of the physical vacuum starting from a first 
principle microscopic theory that describes an elementary quantum 
fluid \cite{laughlin}. Consequently, we have assumed that the 
effective masses of the particles in the elementary quantum fluid 
at zero temperature correspond to the rest masses of the elementary 
particles in the physical vacuum. These effective masses were 
obtained through radiative corrections, at one-loop order, in the 
context of the real time formalism of quantum field theory at finite 
temperature and density. The elementary quantum fluid was described 
in structure and dynamics by the SMWHS and it was characterized by 
non-vanishing chemical potentials associated to the different 
fermion flavour species $(\mu_{fi} \neq 0)$. Subsequently we have 
introduce in the SMWHS a massless neutral scalar field leading to 
Yukawa coupling terms in the Lagrangian density. The effective 
masses have been obtained as function of the coupling constants 
and the unknown fermionic chemical potentials. Starting from the 
experimental mass values for quarks and leptons we have computed, 
as an evidence of the consistency of our approach, the values for 
the $W^{\pm}$ and $Z^0$ boson masses in agreement with the 
experimental values. We have found that if we use the $\mu_{fi}$ 
values obtained from the experimental fermion masses, the gauge 
boson masses are $M_{W} = 81.061 \pm 2.367$ GeV and 
$M_{Z} = 91.917 \pm 2.684$ GeV. For the case $Y=0$, in which the 
scalar boson is absent in the effective model, we have obtained 
from $m_{t} = 172.916$ GeV the experimental central values 
$M_{W} = 80.419$ GeV and $M_{Z} = 91.188$ GeV. For the case 
$0 < Y < 0.607$, we have calculated the experimental mass central 
values for the $W^{\pm}$ and $Z^0$ gauge bosons using a masss value 
for the top quark mass in the range $169.2$ GeV $< m_{t} < 178.6$ 
GeV. This range for the top quark mass implies that the scalar 
boson mass must be in the range $0 < M_{H} < 152$ GeV.

\hspace{3.0mm}

\section*{Acknowledgments} This work was supported by COLCIENCIAS
(Colombia) and by Universidad Nacional de Colombia under research 
grant DIB $803629$. We thank Rafael Hurtado, Yeinzon Rodr\'{\i}guez, 
Marta Losada, Maurizio De Sanctis, Carlos Avila and Germ\'an Sinuco 
by stimulating discussions.

%\section*{References}

\newpage

\begin{figure}[1]
%%Begin InstantTeX Picture
\let\picnaturalsize=N
\def\picsize{3in}
\def\picfilename{Figure1.eps}
\caption{One-loop diagram contributions to the fermionic self-energy}
\label{Fig.(1)}
\end{figure}

\begin{figure}[2]
%%Begin InstantTeX Picture
\let\picnaturalsize=N
\def\picsize{3in}
\def\picfilename{Figure2.eps}
\caption{One-loop diagram contributions to the polarization tensors}
\label{Fig.(2)}
\end{figure}

\begin{figure}[3]
%%Begin InstantTeX Picture
\let\picnaturalsize=N
\def\picsize{3in}
\def\picfilename{Figure3.eps}
\caption{Dependence between the scalar boson and top quark masses}
\label{Fig.(3)}
\end{figure}

\end{document}